\documentstyle[amssymb,prl,aps,epsfig]{revtex}

\begin{document}

\title{Magnetic field induced finite size effect in type-II superconductors}
\author{T. Schneider\\
Physik-Institut der Universit\"{a}t Z\"{u}rich, Winterthurerstrasse 190,\\
CH-8057 Z\"{u}rich, Switzerland} \maketitle
\begin{abstract}
We explore the occurrence of a magnetic field induced finite size
effect on the specific heat and correlation lengths of anisotropic
type-II superconductors near the zero field transition temperature
$T_{c}$. Since near the zero field transition thermal fluctuations
are expected to dominate and with increasing field strength these
fluctuations become one dimensional, whereupon the effect of
fluctuations increases, it appears unavoidable to account for
thermal fluctuations. Invoking the scaling theory of critical
phenomena it is shown that the specific heat data of nearly
optimally doped YBa$_{2}$Cu$_{3}$O$_{7-\delta }$ are inconsistent
with the traditional mean-field and lowest Landau level predictions
of a continuous superconductor to normal state transition along an
upper critical field $H_{c2}\left( T\right) $. On the contrary, we
observe agreement with a magnetic field induced finite size effect,
whereupon even the correlation length longitudinal to the applied
field $H$ cannot grow beyond the limiting magnetic length $L_{H}$
$\propto \sqrt{\Phi _{0}/H}$.  It arises because with increasing
magnetic field the density of vortex lines becomes greater, but this
cannot continue indefinitely. $L_{H}$ is then roughly set on the
proximity of vortex lines by the overlapping of their cores. Thus,
the shift and the rounding of the specific heat peak in an applied
field is traced back to a magnetic field induced finite size effect
in the correlation length longitudinal to the applied field.
\end{abstract}

\bigskip

The superconductor to normal state transition in conventional low
$T_{c}$ materials appears to be well described by the
Ginzburg-Landau mean-field approximation. Because of the large
correlation volume in these materials, the region in which critical
fluctuations are important is too small to be accessible
experimentally. In contrast, with the discovery of superconductivity
in the cuprates, a new era started \cite{bed}. Indeed, marked
deviations from mean-field behavior have been observed over a
temperature range of the order of $10$ K above and below $T_{c}$
\cite{tsda,tshkws,ohl,hub,jacc,kamal,kamal2,pasler,cooper,tsjh,tsjs,jhts,jhts2,tsjshou,book,meingast,tsphysB,osborn,bled,tsdic,parks}.
Theoretical expectations of the kind of critical behavior which
might be observed are: (i) If fluctuations in the vector potential
can be ignored, then the zero-field transition belongs to the
universality class of the three dimensional XY-model, as is the
superfluid transition in $^{4}$He. In an applied magnetic field the
critical behavior is then equivalent to that of uniformly rotating
$^{4}$He near the superfluid transition \cite{parks,vinen}; (ii)
When fluctuations in the vector potential are included \ the charge
of the Cooper pairs, entering via the effective dimensionless charge
$\widetilde{e}=\xi /\lambda =1/\kappa $, charged critical behavior
is expected to occur in which both the correlation length $\xi $ and
the magnetic penetration depth $\lambda $ grow with the same
critical exponent by approaching $T_{c}$ from below
\cite{dasgupta,kleinert,kolnberger,kiometzis,herbut1,herbut2,olsson,hove,sudbo,mo}.
However, in extreme type-II superconductors where $\kappa >>1$ the
effective charge $\widetilde{e}$ is very small. As a consequence the
region close to $T_{c}$, where the system crosses over to the regime
of charged fluctuations, becomes too narrow to access. For instance,
optimally doped YBa$_{2}$Cu$_{3}$O$_{7-\delta }$, while possessing
an extended regime of critical fluctuations, is too strongly type-II
to observe charged critical fluctuations
\cite{tsda,tshkws,ohl,hub,jacc,kamal,kamal2,pasler,cooper,tsjh,tsjs,jhts,jhts2,tsjshou,book,meingast,tsphysB,osborn,bled,tsdic,parks}.
In strongly type-II superconductors ($\kappa \ >>1$) the crossover
upon approaching $T_{c}$ is thus initially to the critical regime of
a weakly charged superfluid where the fluctuations of the order
parameter are essentially those of an uncharged superfluid or
XY-model. Furthermore, there is the inhomogeneity induced finite
size effect which renders the asymptotic critical regime
unattainable \cite{bled,tsdic}. However, underdoped cuprates appear
to open a window onto the charged critical regime because $\kappa $\
becomes rather small in this doping regime. Here the cuprates
undergo a quantum superconductor to insulator transition in the
underdoped limit \cite{book,parks} and correspond to a 2D disordered
bosonic system with long-range coulomb interactions. Close to this
quantum transition $T_{c}$, $\lambda _{ab}$ and $\xi _{ab}$ scale as
$T_{c}\propto \lambda _{ab}^{-2}$ $\propto \xi _{ab}^{-z}$
\cite{book,parks}, yielding with the dynamic critical exponent $z=1$
\cite{book,parks,mpaf,cha,herbutz}, $\kappa _{ab}\propto
\sqrt{T_{c}}$. Recent measurements of the magnetic in-plane
penetration depth of underdoped YBa$_{2}$Cu$_{3}$O$_{6.59}$ clearly
uncovered critical behavior associated with a charged critical
point, in which both the coherence length and the magnetic
penetration depth grow by approaching $T_{c}$ from below with the
same critical exponent \cite{ts123charg}. Thus, as far as static
zero field critical phenomena are concerned, there is little doubt
that near optimum doping the observable critical behavior of bulk
YBa$_{2}$Cu$_{3}$O$_{7-\delta }$ is governed by the three
dimensional(3D) XY universality class. Accordingly, we expect that
the critical behavior in an applied magnetic field is equivalent to
that of uniformly rotating $^{4}$He near the superfluid transition
\cite{parks,vinen}. The singular part of the free energy per unit
volume should then scale as \cite{book,parks}
\begin{equation}
f_{s}=\frac{Q^{\pm }k_{B}T\gamma }{\left( \xi _{ab}^{\pm }\right)
^{3}}G\left( z\right) ,\text{ }z=\frac{H_{c}\left( \xi _{ab}^{\pm
}\right) ^{2}}{\Phi _{0}},\xi _{ab}^{\pm }=\xi _{ab0}^{\pm
}\left\vert t\right\vert ^{-\nu },G^{\pm }\left( 0\right) =1,
\label{eq1}
\end{equation}%
for a magnetic field $H_{c}$ applied parallel to the $c$-axis.
$G\left( z\right) $ is a universal scaling function, $\pm =sgn\left(
t\right) =sgn\left( T/T_{c}-1\right) $, $\xi _{ab}^{\pm }$ the
zero-field in-plane correlation length with critical amplitude $\xi
_{ab0}^{\pm }$, $\gamma =\xi _{ab0}^{\pm }/\xi _{c0}^{\pm }$ the
anisotropy, and $Q^{\pm }$ is \ a universal constant, fixed by the
3D-XY universality class. The fluctuation contribution to the
magnetization $m=-\partial f_{s}/\partial H_{c}$ scales
then as%
\begin{equation}
\frac{m\left( H_{c},T\right) }{TH_{c}^{1/2}}=-\frac{Q^{\pm
}k_{B}\gamma }{\Phi _{0}^{3/2}}z^{-1/2}\frac{dG}{dz},  \label{eq2}
\end{equation}%
while the singular part of the specific heat, $\widetilde{c}\left(
H_{c},T\right) =c\left( H_{c},T\right) \rho /k_{B}\simeq -\partial
^{2}f_{s}/\partial \left\vert t\right\vert ^{2}$ adopts the scaling
form
\begin{equation}
\widetilde{c}\left( H_{c},T\right) =\frac{A^{\pm }}{\alpha
}\left\vert t\right\vert ^{-\alpha }F^{\pm }\left( z\right)
,~\frac{A^{\pm }\left( \xi _{ab0}^{\pm }\right) ^{3}}{\gamma
}=-Q^{\pm }\alpha \left( 1-\alpha \right) \left( 2-\alpha \right)
=\left( R^{\pm }\right) ^{3},~3\nu =2-\alpha ,  \label{eq3}
\end{equation}
where
\begin{equation}
F^{\pm }\left( z\right) =G^{\pm }\left( z\right) -\frac{2\left( 4\nu
+1\right) }{3\left( 3\nu -1\right) }z\frac{dG^{\pm }\left( z\right)
}{dz}+\frac{4\nu }{3\left( 3\nu -1\right) }z^{2}\frac{d^{2}G^{\pm
}\left( z\right) }{dz^{2}}.  \label{eq4}
\end{equation}
$\rho $ denotes the density and $c\left( H_{c},T\right) $ is in
units of $\left( erg/\left( gK\right) \right) $. In the 3D-XY
universality class are the universal quantities $Q^{\pm }$, $R^{\pm
}$, $\nu $ and $\alpha $ given by \cite{peliasetto}
\begin{equation}
R^{+}\simeq 0.36,\text{ }R^{-}\simeq 0.82,\text{ }A^{+}/A^{-}\simeq
1.06,\text{ }\frac{\xi _{ab0,c0}^{-}}{\xi _{ab0,c0}^{+}}\simeq
2.28,\text{ }\nu =\left( 2-\alpha \right) /3\simeq 0.67  \label{eq5}
\end{equation}
In the presence of a sufficiently small magnetic field $H_{c}$, the
specific heat is then expected to have a singular part which
exhibits the scaling behavior
\begin{equation}
\widetilde{c}\left( H_{c},T\right) =\frac{A^{\pm }}{\alpha }\left(
\frac{H_{c}\xi _{ab0}^{2}}{\Phi _{0}}\right) ^{-\alpha /2\nu
}z^{-\alpha /2\nu }F^{\pm }\left( z\right) =\frac{A^{\pm }}{\alpha
}\left( \frac{H_{c}\xi _{ab0}^{2}}{\Phi _{0}}\right) ^{-\alpha /2\nu
}x^{-\alpha }F^{\pm }\left( x^{-1/2\nu }\right) ,\text{
}x=z^{-1/2\nu }.  \label{eq6}
\end{equation}

The magnetization data \cite{hub,cooper} and the zero-field specific
heat measurements of YBa$_{2}$Cu$_{3}$O$_{7-\delta }$
\cite{ohl,pasler,book} agree well with these predictions. In a
nonzero applied field, one can test the scaling form (\ref{eq6}) of
the specific heat by the extent to which data for
$\widetilde{c}\left( H_{c},T\right) H_{c}^{\alpha /2\nu }$ collapse
to a common curve when plotted as a function of $x$. Here, matters
are complicated by the fact that a different kind of scaling
behavior
\begin{equation}
\widetilde{c}\left( H_{c},T\right) =R\left( x_{L}\right) ,\text{
}x_{L}=\frac{T-T_{c2}\left( H_{c}\right) }{\left( TH_{c}\right)
^{2/3}}, \label{eq7}
\end{equation}
is expected when only the lowest Landau level ($L$) is significantly
occupied \cite{bray,thouless}. Here $T-T_{c2}\left( H_{c}\right) $,
or equivalently, $H_{c}=H_{c2}\left( T\right) $ is the upper
critical field of the Ginzburg-Landau theory. Since $\alpha $ in
Eq.(\ref{eq6}) is very small, and $2\nu \approx 4/3$, the two
predictions are rather hard to distinguish \cite{lawrie1}. Some
authors argue that lowest-Landau-level scaling works just as well as
critical-point scaling
\cite{welp,zhou,tesanovic,pierson,pierson2,pierson3,jeandupeux,roulin,junod,roulin2,junod2}.

Theoretically, the scaling form (\ref{eq6}) is an unambiguous
prediction of the theory of critical phenomena and ought to be
observed sufficiently close to the zero-field critical point. On the
other hand, lowest-Landau-level scaling, relies on the assumption
that the correlation length longitudinal to the applied magnetic
field, diverges along the line $T_{c2}\left( H\right) $
\cite{abrikosov}, whereupon a continuous phase transition from the
superconducting to the normal state is predicted to occur.
Accordingly, the behavior of the correlation length longitudinal to
the applied field is essential to verify the lowest Landau level
prediction. In this context it is instructive to rewrite the scaling
variable $z$ (Eq.(\ref{eq1})) in the form
\begin{equation}
z=\frac{H_{c}\xi _{ab}^{2}}{\Phi _{0}}=\frac{\xi
_{ab}^{2}}{aL_{H_{c}}^{2}},\text{ }L_{H_{i}}=\sqrt{\Phi _{0}/\left(
aH_{i}\right) }, \label{eq8}
\end{equation}
with $a\simeq 3.12$\cite{bled}, related to the average distance
between vortex lines \cite{bled,parks,ullah,haussmann,lortz}. The
scaling function $G\left( z\right) $ is then identical to that of a
system with finite extent $L_{H_{c}}$ in the $ab$-plane
\cite{fisher,cardy}. As a consequence, fluctuations which are
transverse to the applied field are stiff and the
corresponding correlation lengths cannot grow beyond%
\begin{equation}
\sqrt{\xi _{i}^{-}\xi _{j}^{-}}=\sqrt{\Phi _{0}/\left( aH_{k}\right)
} =L_{H_{k}},\ i\neq j\neq k.  \label{eq9}
\end{equation}
Hence, the fluctuations of a bulk superconductor in a magnetic field
are longitudinal to the field and for this reason one dimensional,
as noted by Lee and Shenoy \cite{lee}. Noting that fluctuations
become more important with reduced dimensionality, one expects that
the remaining fluctuations, which are longitudinal to the applied
field, remove the mean-field transition at $T_{c2}\left(
H_{c}\right) $. Indeed, thermal fluctuations destroy the ordered
phase in one dimensional systems with short-range interactions.
Furthermore, calculations treating these interactions within the
Hartree approximation \cite{bray,thouless,ullah,ruggeri}, and
generalizations thereof \cite{lawrie1,brezin,hikami}, find that the
correlation length longitudinal to the applied field remains bounded
as well. In this case the correlation length $\xi _{c}^{-}\left(
H_{c},t\right) $ adopts the scaling form
\begin{equation}
\xi _{c}\left( H_{c},t\right) =\xi _{c0}^{\pm }\left\vert
t\right\vert ^{-\nu }S^{\pm }\left( x\right) ,\text{ }x=\left(
\frac{H_{c}\gamma ^{2}\left( \xi _{c0}^{-}\right) ^{2}}{\Phi
_{0}}\right) ^{-1/2\nu }\left\vert t\right\vert .  \label{eq10}
\end{equation}
The scaling function must behave as
\begin{equation}
S^{\pm }\left( x=\pm \infty \right) =1,\text{ \ }S^{\pm }\left(
x\rightarrow \pm 0\right) =s_{0}\left\vert x\right\vert ^{\nu },
\label{eq11}
\end{equation}
so that for $H_{c}\rightarrow 0$, $\xi _{c}\left( H_{c},t\right)
\propto \left\vert t\right\vert ^{-\nu }$ and for $t\rightarrow 0$,
$\xi _{c}\left( H_{c},t\right) \propto \sqrt{\Phi _{0}/H_{c}\text{
}}$. Thus, the divergence of the correlation length $\xi
_{c}^{-}\left( H_{c},t\right) $ is removed and it adopts a maximum
at $T_{p}\left( H_{c}\right) $ $<T_{c}$, yielding the line

\begin{equation}
H_{pc}\left( t\right) =\frac{\Phi _{0}}{x_{p}^{2\nu }\left( \xi
_{ab0}^{-}\right) ^{2}}\left\vert t\right\vert ^{2\nu }.
\label{eq12}
\end{equation}
On the other hand, the specific heat scales according to
Eq.(\ref{eq3}) as
\begin{equation}
\widetilde{c}\left( H_{c},t\right) =\frac{A^{\pm }}{\alpha
}\left\vert t\right\vert ^{-\alpha }F^{\pm }\left( x\right).
\label{eq13}
\end{equation}
In this case the scaling function behaves as
\begin{equation}
F^{\pm }\left( x=\pm \infty \right) =1,\text{ }F^{\pm }\left( x=\pm
0\right) =f_{\pm }\left\vert x\right\vert ^{\alpha },  \label{eq14}
\end{equation}
so that for $H_{c}\rightarrow 0$, $\widetilde{c}\left(
H_{c},t\right) \propto \left\vert t\right\vert ^{-\alpha }$ and for
$t\rightarrow 0$, $\widetilde{c}\left( H_{c},t\right) \propto \left(
\Phi _{0}/H_{c}\right) ^{-\alpha /2\nu }$. Noting that the ratio
$S^{\pm }\left( x\right) /\left( F^{\pm }\left( x\right) \right)
^{\nu /\alpha }$ tends to constant values in the limits
$x\rightarrow 0$ and $x\rightarrow \pm \infty $ the relation between
the $\xi _{c}\left( t,H_{c}\right) $ and the absolute value of the
specific heat $\left\vert \widetilde{c}\left( H_{c},T\right)
\right\vert $,
\begin{equation}
\xi _{c}\left( H_{c},t\right) =\xi _{c0}^{\pm }\left\vert
\frac{A^{\pm }}{\alpha }\right\vert ^{-\nu /\alpha }\left\vert
\widetilde{c}\left( H_{c},t\right) \right\vert ^{\nu /\alpha
}\frac{S^{\pm }\left( x\right) }{\left( F^{\pm }\left( x\right)
\right) ^{\nu /\alpha }}, \label{eq15}
\end{equation}
as obtained from Eqs.(\ref{eq10}) and (\ref{eq13}), reduces in these
limits to
\begin{equation}
\xi _{c}\left( H_{c},t\right) \propto \left\vert \widetilde{c}\left(
H_{c},t\right) \right\vert ^{\nu /\alpha }.  \label{eq16}
\end{equation}
Thus, when this scenario holds true, the specific heat probes
essentially the correlation length longitudinal to the applied
field. As a remnant of the singularity at $T_{c}$ in zero field it
should exhibit a so called finite size effect \cite{fisher,cardy},
resulting in a smooth peak around $T_{p}$ because the correlation
length $\xi _{c}\left( t,H_{c}\right) $ cannot grow beyond $\xi
_{c}\left( t_{p},H_{c}\right) $ (Eq.(\ref{eq12})). Furthermore,
given experimental data for $\widetilde{c}\left( H_{c},t\right) $,
this scenario can be verified. Indeed in the limits
$H_{c}\rightarrow 0$ and $t\rightarrow \pm 0$ the critical behavior
$\left\vert \widetilde{c}\left( H_{c},t\right) \right\vert ^{\nu
/\alpha }\propto \xi _{c0}^{\pm }\left\vert t\right\vert ^{-\nu }$
with $\xi _{c0}^{-}/\xi _{c0}^{+}\simeq 2.28$ (Eqs.(\ref{eq5}))
should hold. This allows to circumvent the aforementioned
difficulties associated with the comparison of scaling functions
with the prediction of the lowest-Landau-level approach. Indeed, if
there is a magnetic field induced finite size effect on the
correlation length longitudinal to the applied field, the transition
is rounded and the assumption of an upper critical field $H_{c2}$ is
not justified.

\begin{figure}[tbp]
\centering
\includegraphics[totalheight=6cm]{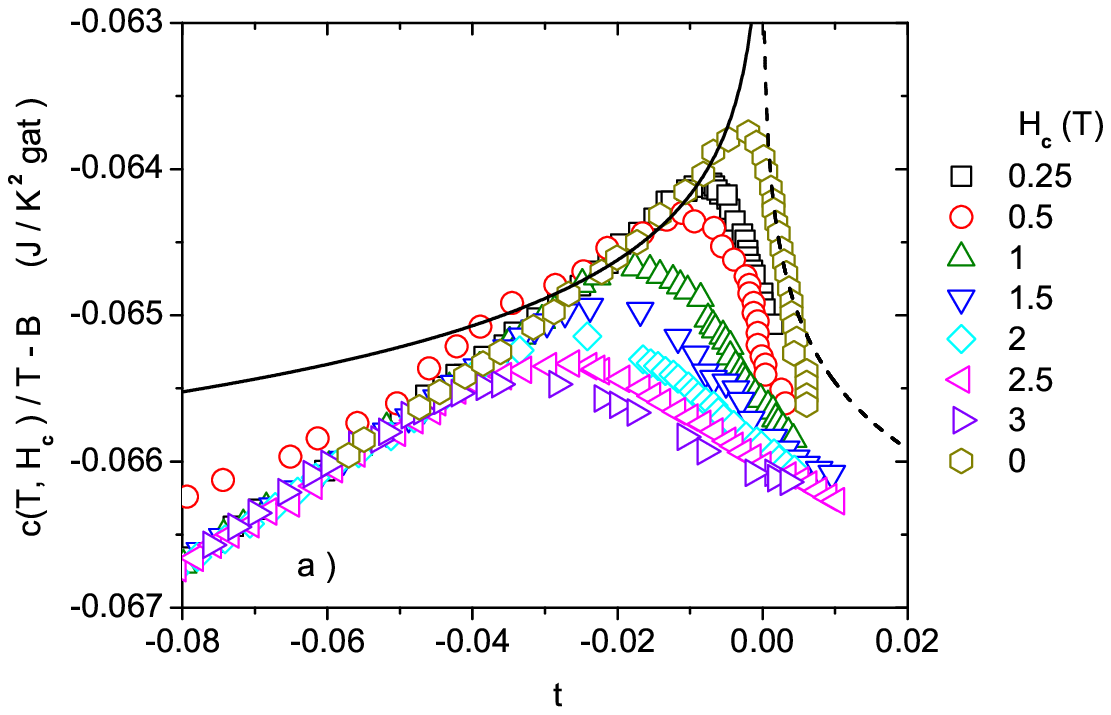}
\includegraphics[totalheight=6cm]{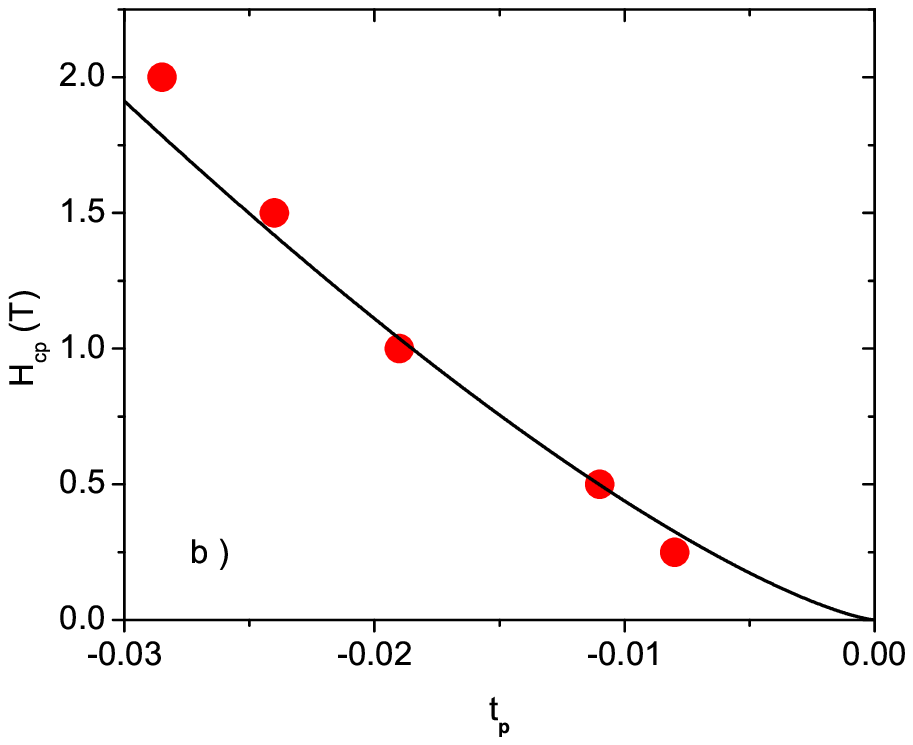}
\caption{a) $\widetilde{c}\left( H_{c},t\right) \propto c\left(
H_{c},T\right) /T-B$ \textit{vs}. $t$ for
YBa$_{2}$Cu$_{3}$O$_{7-\delta }$ derived from Roulin \textit{et al}.
\protect\cite{roulin} with $B=0.1717$ J/K$^{2}$gat \ and
$T_{c}=92.77$ K; The solid and dashed lines are $\widetilde{c}\left(
H_{c},t\right) \propto c\left( H_{c},T\right)
/T-B=\widetilde{A}^{\pm }\left\vert t\right\vert ^{-\alpha }$ with
$\widetilde{A}^{-}=-0.0671$ J/K$^{2}$gat,
$\widetilde{A}^{+}=-0.0684$ J/K$^{2}$gat, and $\alpha =-0.01$. b)
$H_{cp}\left( t\right) $ \textit{vs}. $t$ . The solid line is
$H_{cp}\left( t\right) =210\left\vert t\right\vert ^{2\nu }$ with
$\nu =0.67$, corresponding to Eq.(\ref{eq12}) with $\Phi _{0}/\left(
x_{p}^{2\nu }\left( \xi _{ab0}^{-}\right) ^{2}\right) =210$ (T).}
\label{fig1}
\end{figure}

Here we analyze the specific heat data of Roulin \textit{et al}.
\cite{roulin} to verify the magnetic field induced finite size
scenario. In Fig.\ref{fig1}a we depicted the data for the
YBa$_{2}$Cu$_{3}$O$_{7-\delta }$ single crystal in terms of
$\widetilde{c}\left( H_{c},t\right) \propto c\left( H_{c},T\right)
/T-B$ \textit{vs}. $t=T/T_{c}-1$ with $B=0.1717$ J/K$^{2}$gat \ and
$T_{c}=92.77$ K. The solid and dashed lines are $\widetilde{c}\left(
H_{c},t\right) \propto c\left( H_{c},T\right)
/T-B=\widetilde{A}^{\pm }\left\vert t\right\vert ^{-\alpha }$ with
$\widetilde{A}^{-}=-0.0672$ J/K$^{2}$gat,
$\widetilde{A}^{+}=-0.0685$ J/(K$^{2}$gat), and $\alpha =-0.01$
(Eq.(\ref{eq5})). Using the relations $A^{-}\left( \text{J/(K
gat)}\right) =$ $\widetilde{A}^{-}T_{c}\alpha $, $A^{-}\left(
cm^{-3}\right) =10^{7}/\left( k_{B}V_{gat}\right) A^{-}\left(
\text{J/(K gat)}\right) $, and $V_{gat}=8$ cm$^{3}$ we obtain for
$A^{-}$ the estimate $A^{-}=5.64$ $10^{20}$ cm$^{-3}$ and with the
universal relations (\ref{eq3}) and (\ref{eq5}) for the correlation
volume and the critical amplitudes of the zero field correlation
lengths the estimates
\begin{equation}
\left( \xi _{ab0}^{-}\right) ^{2}\xi _{c0}^{-}\simeq 978\text{ \AA
}^{3},\xi _{ab0}^{-}\simeq 18.43\text{ \AA },\text{ }\xi
_{c0}^{-}\simeq 2.88\text{ \AA ,}  \label{eq17}
\end{equation}
using $\gamma =6.4$. \ As a remnant of the zero-field singularity,
there is for fixed field strength a smeared peak adopting its
maximum at $T_{p}$ which is located below $T_{c}$. As $T_{p}$
approaches $T_{c}$, the peak becomes sharper with decreasing $H_{c}$
and evolves smoothly to the zero-field behavior, smeared by the
inhomogeneity induced finite size effect, arising from the limited
extent $L_{ab,c}$ of the homogeneous domains along the $ab$-plane
and $c$-axes. Since $T_{p}$ decreases systematically with reduced
field down to $H_{c}=0.25$ T, corresponding to $L_{H_{c}}\leq
\sqrt{\Phi _{0}/\left( aH_{c}\right) }=512$ \AA , the magnetic field
sets at and above $H_{c}=0.25$ T the limiting length. On the
contrary, when $L_{H_{c}}\geq L_{ab}$, the inhomogeneities set the
limiting length. In this case, $t_{p}=\left( \xi
_{ab0}^{-}/L_{ab}\right) ^{1/\nu }$, independent of the applied
field. Thus, the field dependence of $t_{p}$ is a characteristic
feature of a magnetic field induced limiting length. The line
$H_{cp}\left( t\right) $ is shown in Fig.\ref{fig1}b. The solid line
is Eq.(\ref{eq12}) with $\Phi _{0}/\left( x_{p}^{2\nu }\left( \xi
_{ab0}^{-}\right) ^{2}\right) =210$ T, yielding with $\xi
_{ab0}^{-}\simeq 18.43$ \AA\ (Eq.(\ref{eq17})),
\begin{equation} x_{p}\simeq 2.21,  \label{eq17b}
\end{equation}
and $z_{p}=x_{p}^{-1/2\nu }\simeq 0.55$, which agrees well with the
previous estimate $z_{p}\simeq a^{-1/2}=3.12^{-1/2}\simeq 0.57$
\cite{bled}.

\begin{figure}[tbp]
\centering
\includegraphics[totalheight=6cm]{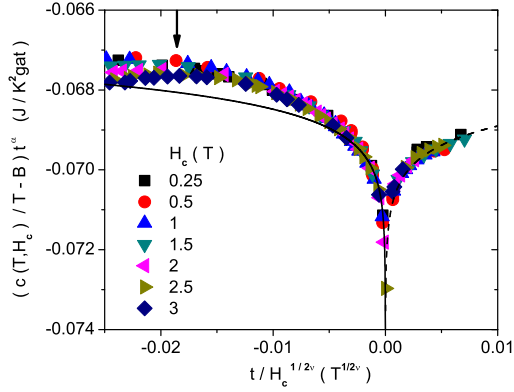}
\caption{$\widetilde{c}\left( H_{c},t\right) \left\vert t\right\vert
^{-\alpha }\propto \left( c\left( H_{c},T\right) /T-B\right)
\left\vert t\right\vert ^{\alpha }$ \textit{vs}. $t/H_{c}^{1/2\nu }$
derived from the data of Roulin \textit{et al}.
\protect\cite{roulin} shown in Fig.\ref{fig1}a. The solid and dashed
lines are $-0.0654\left( -t/H_{c}^{1/2\nu }\right) ^{\alpha }$ and
$-0.0658\left( -t/H_{c}^{1/2\nu }\right) ^{\alpha }$, respectively.
The arrow marks $t_{p}/H^{1/2\nu }\simeq -0.0184$. } \label{fig2}
\end{figure}

In Fig.\ref{fig2} we displayed the scaling plot $\widetilde{c}\left(
H_{c},t\right) \left\vert t\right\vert ^{\alpha }\propto \left(
c\left( H_{c},T\right) /T-B\right) \left\vert t\right\vert ^{\alpha
}$ \textit{vs}. $t/H_{c}^{1/2\nu }$ derived from the data of Roulin
\textit{et al}. \cite{roulin} shown in Fig.\ref{fig1}a. Noting that
according to Eq.(\ref{eq13}), $\left( c\left( H_{c},T\right)
/T-B\right) \left\vert t\right\vert ^{\alpha }\propto
\widetilde{c}\left( H_{c},t\right) \left\vert t\right\vert ^{-\alpha
}=A^{\pm }F^{\pm }\left( x\right) /\alpha $, where $x\propto
t/H_{c}^{1/2\nu }$, this plot uncovers essentially the scaling
function $F^{\pm }\left( x\right) $, whereupon the data points
should fall on a single curve, as they apparently do, when plotted
versus $x\propto t/H_{c}^{1/2\nu } $. The solid and dashed curves
indicate the asymptotic behavior in the limits $x\propto
t/H_{c}^{1/2\nu }\rightarrow \pm 0$ (Eq.(\ref{eq14})), while the
arrow marks $t_{p}/H_{c}^{1/2\nu }=x_{p}\left( \left( \xi
_{ab0}^{-}\right) ^{2}/\Phi _{0}\right) ^{1/2\nu }$, where the peak
in the specific heat adopts its maximum value. However, as
aforementioned, it is difficult to distinguish different models on
the basis of such scaling functions.

\begin{figure}[tbp]
\centering
\includegraphics[totalheight=6cm]{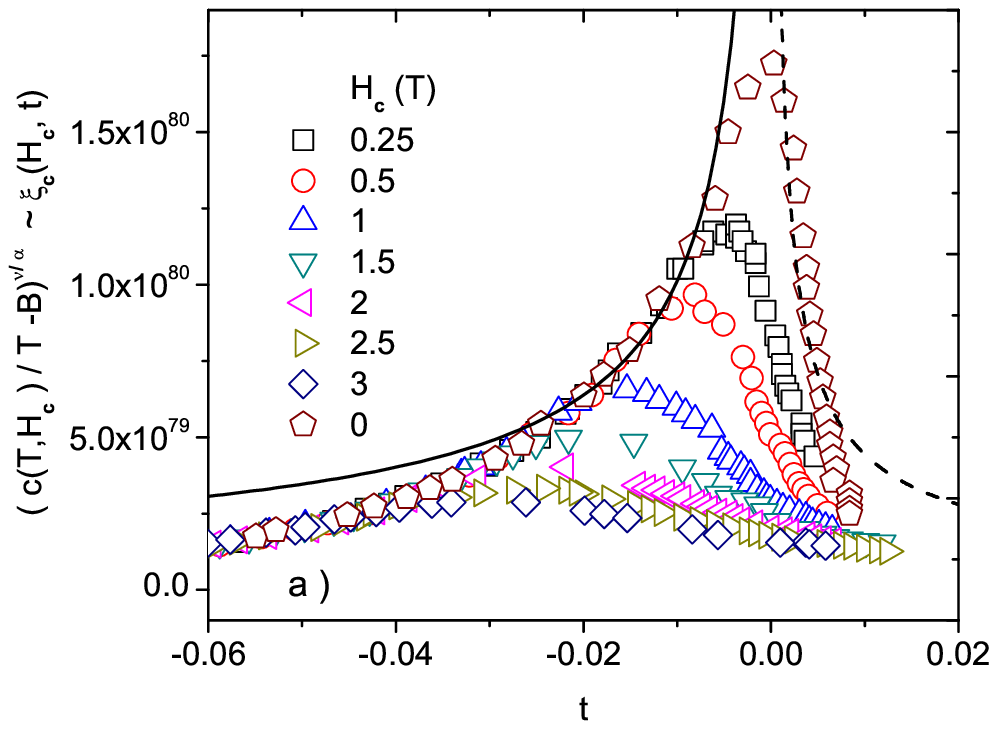}
\includegraphics[totalheight=6cm]{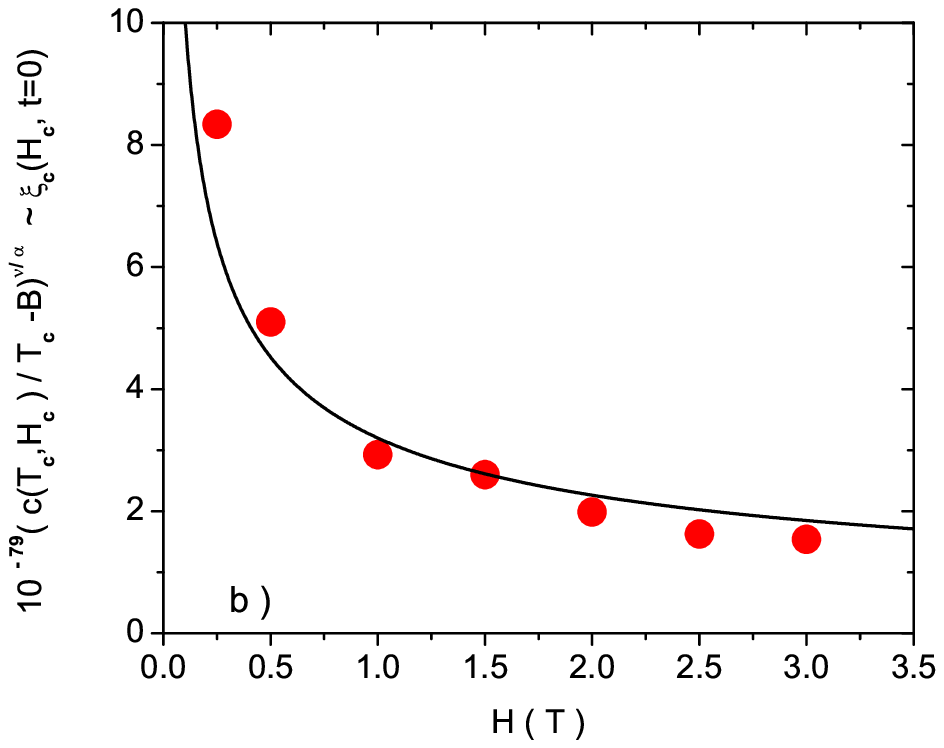}
\caption{a) $\left\vert \widetilde{c}\left( H_{c},t\right)
\right\vert ^{\nu /\alpha }\propto \left\vert c\left( H_{c},T\right)
/T-B\right\vert ^{\nu /\alpha }$ \textit{vs}. $t$ derived from the
data of Roulin \textit{et al}. \protect\cite{roulin} shown in
Fig.\ref{fig1}a. The solid and dashed lines are $\left\vert c\left(
H_{c},T\right) /T-B\right\vert ^{\nu /\alpha }=\Gamma ^{\pm
}\left\vert t\right\vert ^{-\nu }$ with $\Gamma ^{-}=4.7$ 10$^{78}$
, $\Gamma ^{+}=\Gamma ^{-}/2.28$ , and $\nu =0.67$. b) $\left\vert
\widetilde{c}\left( H_{c},t=0\right) \right\vert ^{\nu /\alpha
}\propto \left\vert c\left( H_{c},T_{c}\right) /T_{c}-B\right\vert
^{\nu /\alpha }$ \textit{vs}. $H_{c}$ derived from the data shown in
Fig.\ref{fig3}a. The solid line is $ \left\vert c\left(
H_{c},T_{c}\right) /T_{c}-B\right\vert ^{\nu /\alpha }=3.2 $
$10^{79}H_{c}^{-1/2}$.} \label{fig3}
\end{figure}

However, in view of Eqs.(\ref{eq15}) and (\ref{eq16}), giving the
relationship between the fluctuation contribution to the specific
heat and the correlation length longitudinal to the applied field,
$\xi _{c}\left( t,H_{c}\right) $, the magnetic field induced finite
size scenario can be verified by considering the plot $\left\vert
\widetilde{c}\left( H_{c},t\right) \right\vert ^{\nu /\alpha
}\propto \left\vert c\left( T,H_{c}\right) /T-B\right\vert ^{\nu
/\alpha }$ \textit{vs}. $t$, predicted to probe $\xi _{c}\left(
t,H_{c}\right) $. In Fig.\ref{fig3} we show $\left\vert
\widetilde{c}\left( H_{c},t\right) \right\vert ^{\nu /\alpha
}\propto \left\vert c\left( T,H_{c}\right) /T-B\right\vert ^{\nu
/\alpha }$ \textit{vs}. $t$ derived from the data of Roulin
\textit{et al}. \cite{roulin} shown in Fig.\ref{fig1}a. The solid
and dashed line mark the leading zero field critical behavior in
terms of $\left\vert \widetilde{c}\left( H_{c},t\right) \right\vert
^{\nu /\alpha }\propto \left\vert c\left( T,H_{c}\right)
/T-B\right\vert ^{\nu /\alpha }=\Gamma ^{\pm }\left\vert
t\right\vert ^{-\nu }$, with $\Gamma ^{-}/\Gamma ^{+}=2.28$,
consistent with the universal ratio $\xi _{c0}^{-}/\xi
_{c0}^{+}\simeq 2.28$ of the 3D-XY universality class
(Eq.(\ref{eq5})). This confirms that in the scaling regime
considered here $\left\vert \widetilde{c}\left( H_{c},t\right)
\right\vert ^{\nu /\alpha }$ probes essentially the correlation
length longitudinal to the applied magnetic field, $\xi _{c}\left(
t,H_{c}\right) $, so that Eq.(\ref{eq16}) applies. The rounded peak
in zero field reveals then an inhomogeneity induced finite size
effect, while the smeared peak in finite fields, its shift and
broadening with increasing field strength discloses the magnetic
field induced finite size effect on the correlation length
longitudinal to the applied field. Because $\xi _{c}\left(
t,H_{c}\right) \propto \left\vert \widetilde{c}\left( H_{c},t\right)
\right\vert ^{\nu /\alpha }\propto \left\vert c\left( T,H_{c}\right)
/T-B\right\vert ^{\nu /\alpha }$ the field dependence of $t_{p}$,
where the correlation length adopts its maximum value set by the
magnetic field, coincides with the $t_{p}\left( H_{c}\right) $,
where the specific heat reaches its maximum value. In this context
it is important to recognize that the magnetic field dependence of
$t_{p}$ is a unique consequence of the magnetic field induced finite
size effect. Indeed, when inhomogeneities set the limiting length
then $t_{p}=\left( \xi _{ab0}^{-}/L_{ab}\right) ^{1/\nu } $,
independent of the applied field.

When the magnetic field induced finite size effect scenario is
correct, the occurrence of the effect is not restricted to
temperatures below $T_{c}$. \ It is particularly dramatic at
$T_{c}$, where in a homogeneous system in zero field the correlation
lengths are infinite. In an applied field the scaling form
(\ref{eq10}) yields the prediction
\begin{equation}
\xi _{c}\left( H_{c},t=0\right) =\frac{s_{0}}{\gamma }\left(
\frac{\Phi _{0}}{H_{c}}\right) ^{1/2}=\frac{s_{0}a^{1/2}}{\gamma
}L_{H_{c}}, \label{eq18}
\end{equation}
where we used the definition (\ref{eq8} for the limiting magnetic
length $L_{H_{c}}$. In Fig.\ref{fig3}b we displayed $\xi _{c}\left(
H_{c},t=0\right) \propto \left\vert \widetilde{c}\left(
H_{c},t=0\right) \right\vert ^{\nu /\alpha }\propto \left\vert
c\left( H_{c},T_{c}\right) /T_{c}-B\right\vert ^{\nu /\alpha }$
\textit{vs}. $H_{c}$. The consistency with the solid line, which is
Eq.(\ref{eq18}) in the form $\xi _{c}\left( H_{c},t=0\right) \propto
\left\vert \widetilde{c}\left( H_{c},t=0\right) \right\vert ^{\nu
/\alpha \propto }\left\vert c\left( H_{c},T_{c}\right)
/T_{c}-B\right\vert ^{\nu /\alpha }=3.210^{79}H_{c}^{-1/2}$, reveals
again that an applied magnetic field leads to a finite size effect
in the correlation length longitudinal to the field.

\begin{figure}[tbp]
\centering
\includegraphics[totalheight=6cm]{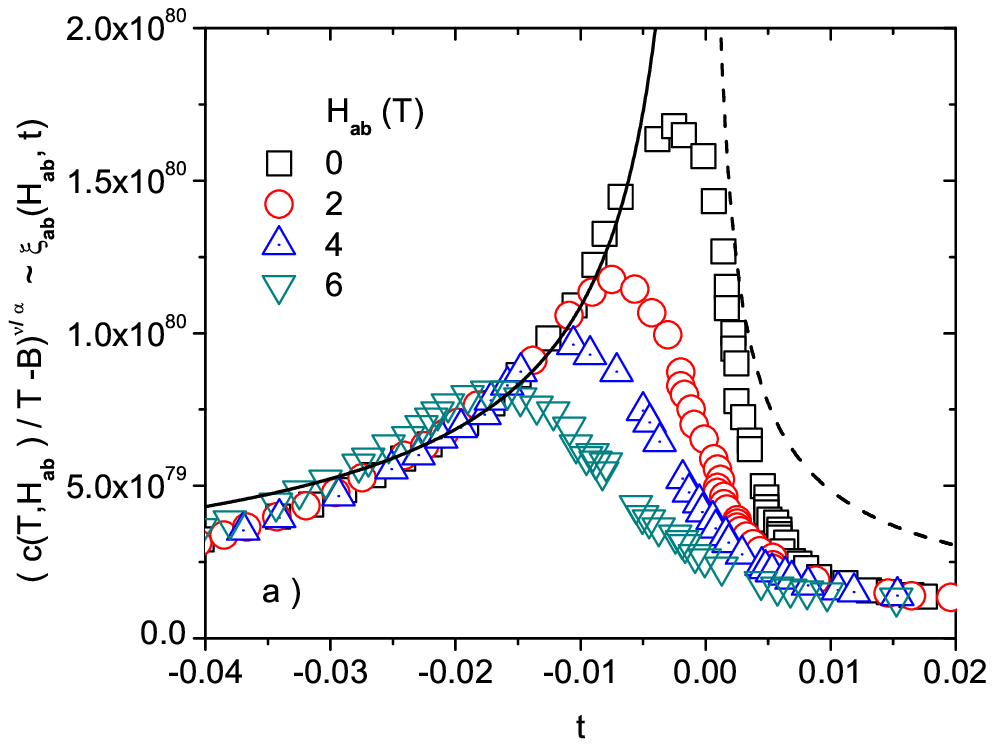}
\includegraphics[totalheight=6cm]{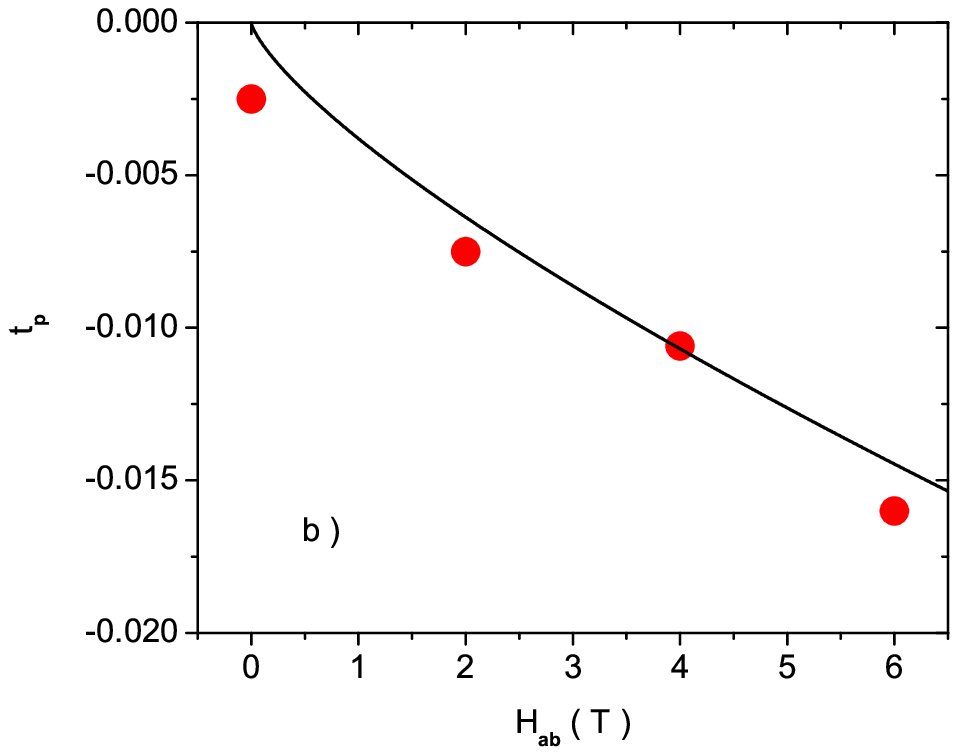}
\caption{a) $\left\vert \widetilde{c}\left( H_{ab},t\right)
\right\vert ^{\nu /\alpha }\propto \left\vert c\left(
H_{ab},T\right) /T-B\right\vert ^{\nu /\alpha }$ \textit{vs}. $t$
derived from the data of Roulin \textit{et al}.
\protect\cite{roulin} with $B=0.1717$ J/K$^{2}$gat \ and
$T_{c}=92.77$ K. The solid and dashed lines are $\left\vert c\left(
H_{ab},T\right) /T-B\right\vert ^{\nu /\alpha }=\Gamma ^{\pm
}\left\vert t\right\vert ^{-\nu }$ with $\Gamma ^{-}=5.05$ 10$^{78}$
, $\Gamma ^{+}=\Gamma ^{-}/2.28$ , and $\nu =0.67$. b) $t_{p}$
\textit{vs}. $H_{ab}$. The solid line is
$t_{p}=-0.0034H_{ab}^{1/2\nu }$, which corresponds to
Eq.(\ref{eq20}) with $\left( \left( x_{p}^{2\nu }\xi _{ab0}^{-}\xi
_{c0}^{-}\right) /\Phi _{0}\right) ^{1/2\nu }=0.0034$ T.}
\label{fig4}
\end{figure}

For fields applied parallel to the $a$-axis, the transverse
correlation lengths $\xi _{b}$ and $\xi _{c}$ are according to
Eq.(\ref{eq8}) bounded by $\sqrt{\xi _{a}^{-}\xi
_{b}^{-}}=L_{H_{a}}$. When the magnetic field induced finite size
scenario holds true, the correlation length longitudinal to the
applied field, $\xi _{a}\left( H_{a},t\right) $, should be bounded
as well. In analogy to Eq.(\ref{eq10}) the longitudinal correlation
length adopts then the scaling form
\begin{equation}
\xi _{a}\left( H_{a},t\right) =\xi _{a0}^{\pm }\left\vert
t\right\vert ^{-\nu }S^{\pm }\left( x\right) ,\text{ }x=\left(
\frac{H_{a}\xi _{b0}^{-}\xi _{c0}^{-}}{\Phi _{0}}\right) ^{-1/2\nu
}\left\vert t\right\vert ,  \label{eq19}
\end{equation}
with the limiting behavior given in Eq.(\ref{eq11}). Thus, the
divergence of the correlation length is removed and $\xi
_{a}^{-}\left( H_{a},t\right) $ adopts at $T_{p}\left( H_{a}\right)
$ $<T_{c}$ a maximum, yielding the line

\begin{equation}
H_{pa}\left( t\right) =\frac{\Phi _{0}}{x_{p}^{2\nu }\xi
_{b0}^{-}\xi _{c0}^{-}}\left\vert t\right\vert ^{2\nu }.
\label{eq20}
\end{equation}
Furthermore, in analogy to Eq.(\ref{eq16}) the relation
\begin{equation}
\xi _{a}\left( H_{a},t\right) \propto \left\vert \widetilde{c}\left(
H_{a},t\right) \right\vert ^{\nu /\alpha },  \label{eq21}
\end{equation}
between the longitudinal correlation length and the specific heat
should hold in the limits $x\rightarrow 0$ and $x\rightarrow \pm
\infty $. In Fig.\ref{fig4}a we depicted $\left\vert
\widetilde{c}\left( H_{ab},t\right) \right\vert ^{\nu /\alpha
}\propto \left\vert c\left( H_{ab},T\right) /T-B\right\vert ^{\nu
/\alpha }$ \textit{vs}. $t$ derived from the data of Roulin
\textit{et al}. \cite{roulin}. The solid and dashed line mark the
leading zero field critical behavior in terms of $\left\vert
\widetilde{c}\left( H_{c},t\right) \right\vert ^{\nu /\alpha
}\propto \left\vert c\left( T,H_{c}\right) /T-B\right\vert ^{\nu
/\alpha }=\Gamma ^{\pm }\left\vert t\right\vert ^{-\nu }$, with
$\Gamma ^{-}/\Gamma ^{+}=2.28$, consistent with the universal ratio
$\xi _{ab0}^{-}/\xi _{ab0}^{+}\simeq 2.28$ of the 3D-XY universality
class (Eq.(\ref{eq5})). This consistency confirms again that in the
scaling regime considered here $\left\vert \widetilde{c}\left(
H_{ab},t\right) \right\vert ^{\nu /\alpha }$ probes the correlation
length $\xi _{ab}\left( t,H_{ab}\right) $ longitudinal to the
applied field. The rounded peak in zero field reveals an
inhomogeneity induced finite size effect, while the smeared peak in
finite fields, its shift and broadening with increasing field
strength disclose the magnetic field induced finite size effect in
$\xi _{ab}\left( t,H_{c}\right) $. Indeed, from Fig.\ref{fig4}b it
is seen that the field dependence of $t_{p}$ where the correlation
length $\xi _{ab}\left( t,H_{c}\right) $ adopts its maximum value,
set by the magnetic field, is consistent with
$t_{p}=-0.0034H_{ab}^{1/2\nu }$ which is Eq.(\ref{eq20}) with
$\left( \left( x_{p}^{2\nu }\xi _{ab0}^{-}\xi _{c0}^{-}\right) /\Phi
_{0}\right) ^{1/2\nu }=0.0034$ (T), resulting from the $\xi
_{ab,c0}^{-}$ given by Eq.(\ref{eq17}) and $x_{p}=2.21$
(Eq.(\ref{eq17b}).

To summarize our result for an anisotropic type-II superconductor,
we have shown that near the zero field transition temperature
superconductivity is in a magnetic field subjected to a field
induced finite size effect. The crucial ingredient for a finite size
effect is an energy gap in the excitation spectrum of fluctuations.
In the present case it is the discrete set of Landau levels. Indeed,
there is the formal analogy with the Landau levels of a charged
particle moving in circular orbits in the plane perpendicular to the
applied field at the cyclotron frequency. As a consequence, the
fluctuations which are transverse to the field are stiff and have a
length scale $L_{H}\propto \sqrt{\Phi _{0}/H}$. Hence, the
fluctuations of a bulk type-II superconductor become one dimensional
and are longitudinal to the applied field, as noted by Lee and
Shenoy \cite{lee}. Because fluctuations become more important with
reduced dimensionality, one expects then that the interaction of
these fluctuations remove the mean-field transition at $T_{c2}\left(
H_{c}\right) $, because thermal fluctuations destroy the ordered
phase in one dimensional systems with short-range interactions. The
absence of this transition is further supported by calculations
treating the fluctuations within the Hartree approximation
\cite{bray,thouless,ullah,ruggeri}, and generalizations thereof
\cite{lawrie1,brezin,hikami}. They suggest that the correlation
length longitudinal to the applied field remains bounded as well.
Invoking the scaling theory of critical phenomena we confirmed this
prediction. We have shown that the specific heat data of Roulin
\textit{et al}. \cite{roulin2} clearly reveals a magnetic field
induced finite size effect in the correlation length longitudinal to
the applied field. Accordingly, there is no evidence for a phase
transition line $T_{c2}\left( H\right) $ near the zero field
transition temperature $T_{c}$.

\acknowledgments I would like to thank K. A. M\"{u}ller for
stimulating discussions on this and related subjects.

\end{document}